\documentclass[12pt]{aastex}
\usepackage{psfig}
\usepackage{emulateapj5}
\newcommand{\aips}{{\sc AIPS}}
\newcommand{\miriad}{{\sc MIRIAD}}
\newcommand{\mjyb}{mJy beam${}^{-1}$}
\begin{document}

\title{A Wide Field, Low Frequency Radio Survey of the Field of M31: 
I. Construction and Statistical Analysis of the Source Catalog} 
\author{Joseph D. Gelfand}
\affil{Harvard-Smithsonian Center for Astrophysics}
\affil{60 Garden St. MS-10 Cambridge, MA 02138}
\email{jgelfand@cfa.harvard.edu}
\and
\author{T. Joseph W. Lazio}
\affil{Naval Research Laboratory - Code 7213}
\affil{4555 Overlook Ave. SW Washington, DC 20375-5351}
\email{Joseph.Lazio@nrl.navy.mil}
\and
\author{B. M. Gaensler}
\affil{Harvard-Smithsonian Center for Astrophysics}
\affil{60 Garden St. Cambridge, MA 02138}
\email{bgaensler@cfa.harvard.edu}

\begin{abstract}
We present here the results of a 325~MHz radio survey of M31, conducted with 
the A-configuration of the Very Large Array.  The survey covered an area of 
7.6~deg$^2$, and a total of 
405 radio sources between $\la$6\arcsec~and 170\arcsec~in extent were mapped 
with a resolution of 6\arcsec~and a 1$\sigma$~sensitivity of $\sim$0.6~\mjyb.
For each source, its morphological class, major axis $\theta_M$, minor axis 
$\theta_m$, position angle $\theta_{PA}$, peak flux $I$, integrated flux 
density $S$, spectral index $\alpha$ and spectral
 curvature parameter $\varphi$ were calculated.   A comparison of the flux and 
radial distribution -- both in the plane of the sky and in the plane of M31 -- 
of these sources with those of the XMM--LSS and WENSS radio surveys revealed 
that a vast majority of sources detected are background
 radio galaxies.  As a result of this analysis, we expect that only a few 
sources are intrinsic to M31.  These sources are identified and 
discussed in an accompanying paper.
\end{abstract}

\section[Intro]{Introduction}
In order to understand the dynamics of a galaxy it is crucial 
to know properties of its radio population.  Much has been learned about the 
Milky Way from radio observations of the supernova remnants (SNRs), 
H{\sc II} regions, and pulsars detected within it.  However, a complete census 
of radio sources in the Milky Way is difficult to obtain because of source 
confusion.  In addition, since the distances to these sources are often 
extremely uncertain, it is difficult to determine their properties.  As 
a result, it is advantageous to observe external galaxies to learn about the 
dynamics and properties of ``normal'' galaxies.  The first step of this process
 is to obtain a census of radio populations in a galaxy, and it is for this 
reason that we have surveyed the radio population of M31, the nearest spiral 
galaxy.

M31 has been surveyed in the radio before, both as parts of larger surveys 
such as the WENSS \citep{wenss} and NVSS \citep{nvss} surveys, and as the 
focus of dedicated surveys such as the 36W \citep{36w}, 37W \citep{37w}, and 
Braun \citep{braun} surveys, the properties of all of which are summarized in 
Table \ref{catprop}.  Despite its proximity to the Milky Way, mapping M31 is 
difficult because its large angular size on the sky ($>$2$^\circ$) requires 
many pointings at higher radio frequencies ($\nu \ga 1$~GHz) to fully cover.  
As a result, existing surveys of M31 are either deep but cover only a small 
region of M31, or cover the entire optical disk of M31 with relatively poor 
sensitivity, as shown in Table \ref{catprop} and Figure \ref{pbeam}.  
Because of this, the radio population of discrete sources in M31 is not well 
understood.  To rectify this situation, we have surveyed M31 with the Very 
Large Array\footnote{The VLA is operated by The National Radio Astronomy 
Observatory, which is a facility of the National Science Foundation operated 
under cooperative agreement by Associated Universities, Inc.} (VLA) at 325~MHz
($\lambda=90~\mbox{cm}$) using the A-configuration, achieving a 1$\sigma$ 
sensitivity less than a mJy with a resolution of 6\arcsec~($\sim$20 pc at the 
distance of M31, assumed to be 780~kpc; \citeauthor{stanek} \citeyear{stanek}) 
over the entire optical disk of M31 because of the large size of the VLA's 
primary beam at this frequency.  Using the A-configuration has the advantage 
that extended emission from M31 was resolved out, allowing us to better 
determine the properties of the compact radio population.  As seen in Table 
\ref{catprop}, the survey presented here does very well in resolution 
($\theta_{res}$) vs. field of view (FOV) when compared to past surveys of M31, 
and has a similar sensitivity to that of previous higher frequency surveys.

This paper describes the observations, the data reduction process that led 
 to the final source list, and the statistical properties of the detected 
sources.  This paper is structured as follows: Section \ref{analysis} describes
 the observations and data reduction techniques used to make the final 
source list, Section \ref{sourcelist} presents the final (``GLG'') source 
list, Section \ref{props} presents the statistical properties of the GLG 
source list, and Section \ref{conclusions} presents the conclusions derived 
from this analysis.  In a companion paper \citep{paper2}, we will classify the 
sources and discuss their properties.  

This paper uses the convention of $S_{\nu}$ is the flux density of a source at 
a frequency $\nu$ MHz, radio spectral index $\alpha$ is defined as 
$S_{\nu} \propto \nu^{\alpha}$, and the distance to M31 is assumed to be 
780$\pm$13~kpc \citep{stanek}.

\section[Analysis]{Observations and Data Reduction}
\label{analysis}
This section describes the observations and data reduction process that 
led to the GLG source list.  As detailed below, the data reduction process 
differs from that of higher frequency observations because of the large field 
of view.  In addition, 325~MHz observations are plagued by radio frequency 
interference (RFI) which must be removed before imaging.

\subsection{Observations}
\label{observations}
This paper is the result of a five hour ($\sim$4 hours on-source) observation 
of M31 conducted on 2000 December 15 with the VLA A-array.  The observations 
were done in spectral line mode with two intermediate frequencies (IFs), 
centered at 321.6~MHz and 327.5~MHz, each with a bandwidth 
of 3125 kHz divided into 64 frequency channels.  Spectral line mode was used to
 reduce 
the effect of primary beam chromatic aberration (bandwidth smearing) and, more 
importantly, to expedite excision of radio frequency interference (RFI).  The 
receivers recorded both Right Circular Polarization (RCP) and Left Circular 
Polarization (LCP) data, which means that only Stokes I (total intensity) and 
Stokes V (circular polarization) were measured.  Calibrator sources 3C~405 
(Cyg A) and TXS 0035+413 were observed hourly for $\sim$5 minutes. Cyg A was 
used to set the flux density scale and calibrate the bandpass by applying a 
333 MHz model to the observed image.\footnote{Available at 
http://rsd-www.nrl.navy.mil/7213/lazio/tutorial/VLAmodels/models.html)}  
This model assumes that Cyg A has a total flux density of 5983 Jy 
at 325 MHz, within the 6\% error of the Baars flux density scale \citep{baars}.
  The calibrator source TX 0035+413 was then used to calibrate the 
visibilities.  
Table \ref{obsprop} summarizes the properties of the observation.\footnote{The 
FOV listed in this table is equal to $FOV=\pi R^{2}$, where $R$ is the distance
 in degrees from the pointing center to the most distant source detected.  This
 is larger than the FWHM of the primary beam due to the shapes of the 
``facets'' discussed in Section \ref{imgprod}}

\subsection{Data Reduction}
\label{datared}
The initial phase and flux density calibration was done using the \aips~task 
{\sc CALIB}.  For both Cyg A and TXS 0035+413, edge channels 1-5 and 
58-63 were flagged due to poor frequency response and the visibility data was 
inspected manually for RFI.  During the first half of the first scan of Cyg~A, 
antenna 13 malfunctioned.  To preserve data from this antenna taken in the 
second half of this scan, all data points involving antenna 13 in the 
first-half of this scan were flagged and the phase and flux solutions 
calculated for the second half were renormalized.  After the flux and phase 
calibration solutions were calculated, they were applied to all the visibility 
data, including the M31 visibilities.

The process of converting the raw data into a usable image and source list was
a long and complicated one, as illustrated in Figure \ref{flwchrt}.  The major 
complication in this process is the presence of RFI, and it was necessary to 
remove as much of it as possible before any imaging.  This was accomplished by 
inspecting each baseline visually for RFI.  Both the RCP and LCP data were 
searched separately for RFI because, being terrestrial in origin, it was often 
polarized and would appear in one polarization channel but not the other.  To 
avoid complications when calculating the total intensity (Stokes I), flags in 
one polarization were applied to the other.  Since polarized RFI is strong in 
Stokes V (Stokes V=RCP--LCP), all baselines were searched for RFI in this 
Stokes parameter as well.

\subsubsection{Image Production}
\label{imgprod}
Only after all the RFI was removed was it possible to produce a high quality 
image of M31.  Due to its large FOV at 325~MHz, the VLA is not coplanar at 
this frequency.  Therefore, in order to make the image using standard radio 
techniques, the primary beam was divided into 73 small ``facets''
over which the VLA can safely assumed to be coplanar \citep{flatn}.

Two sets of images were made, both created by an iterative process of 
CLEANing and self--calibration.  For Images A, the self--calibration process 
used every positive CLEAN component (CC). As a result, the calibration model 
included CCs that were peaks in the noise, artificially lowering the noise 
of the image and increasing the probability of a false detection. However, 
these maps were good enough to find sources (Source list I) that generated the 
restrictive clean boxes (CLEAN Boxes 2) used to image Images B.

The second set of images (Images B), which generated the source list presented 
in this paper, were also created using an iterative process involving the 
CLEAN algorithm and self--calibration.  However, this time CCs were only placed
 around sources in Source list I and off-axis WENSS sources (CLEAN Boxes 2). 
The use of smaller clean boxes decreased the possibility of ``over-cleaning'' 
and removes any CLEAN bias (see \citeauthor{nvss} \citeyear{nvss} for a good 
description of CLEAN 
bias).  In addition, the self--calibration process used only the strongest CCs 
inside the primary beam to maximize the accuracy of the phase calibration model
 inside the primary beam.  Using this method, we generated maps with a typical 
noise level of $\sim$0.6 mJy beam${}^{-1}$ and a signal-to-noise ratio of 
$\sim$1000 around the strongest sources -- equivalent to that of the highest 
quality low--frequency radio images.

\subsubsection{Generating the Source list}
\label{gensrc}
Sources in the first (Images A) and second (Images B) set of images 
were found using the \miriad~task {\sc SFIND} v2.0, which
uses a false discovery rate (FDR) algorithm to detect sources \citep{sfind}.  
This algorithm is similar to a Gaussian source-finding algorithm, except 
that it also accounts for noise variations on a user-defined size (the 
{\it rmsbox} parameter).  For each one of the central 73 facets, we made six 
source lists.   The first three source list were made using {\it rmsbox}=1.25, 
1.88, and 2.5 beams (20, 30, and 40 pixels).  Multiple values were used for 
the {\it rmsbox} parameter because sources the size of {\it rmsbox} were often 
missed by {\sc SFIND}, and values of {\it rmsbox} larger than 2.5 beams 
caused {\sc SFIND} to miss many sources.  For the different values of 
{\it rmsbox}, {\sc SFIND} detected a similar number of sources but there 
were sources found for one value but not others.  These three source lists 
were 
then compiled into one list. Since the facets overlapped slightly, sources on 
the edge of a facet were ``discovered'' more than once.  When merging the 
source list of each facet into a single source list, the source with the 
strongest signal-to-noise from this set of duplicates was kept.  
{\sc SFIND} also had difficulty with large and/or complicated sources, 
the integrated properties of these sources were determined by running 
{\sc SFIND} without its FDR capabilities\footnote{Without the FDR 
algorithm, {\sc SFIND} works like other Gaussian source finding programs 
-- looking for regions with a signal-to-noise greater than a user-defined 
value. In our case, we instructed {\sc SFIND} to keep all 
$\geq$5$\sigma$ sources.} on maps smoothed to a resolution of 
20\arcsec$\times$20\arcsec.  In these smoothed maps, a complicated source 
(e.g. a resolved radio jet) would be unresolved, allowing {\sc SFIND} to 
better determine its integrated properties.  As with the high resolution maps, 
three source lists were made, each with a different value of {\it rmsbox}, 
{\it rmsbox}=3.13, 6.25, and 9.38 beams (50, 100, and 150 pixels).  Again, 
these three source lists were compiled into one.  Once the final 
high--resolution and smoother source lists were made, they were compiled into 
one.  In the case where a source in the smoothed map overlapped many sources 
in the high-resolution maps, the smoothed source was kept.  Sources that 
appeared only in the smoothed maps were kept as well.  However, in all other 
cases the source in the high-resolution maps were kept.  The positions and 
sizes of sources found in Images A (Source list I) defined the clean boxes 
(CLEAN Boxes 2) used to make the second and final set of images (Images B).  
The final source list presented here (hereafter referred to as the GLG source 
list), given in Table \ref{srclist}, was made from Images B using this 
procedure.  The location of these sources is shown in Figure \ref{m31_325}.

\subsection{Source Properties}
\label{srcprop}
The \miriad~task {\sc SFIND} determines the RA, DEC, local flux density 
RMS $\sigma_{RMS}$, peak brightness $I$, integrated flux density $S$, image 
major axis $\theta_M$, image minor axis $\theta_m$, and position angle 
$\theta_{PA}$ of every source it finds.  Sources were separated into four 
morphological categories -- unresolved (``U''), elongated (``El''), complex 
(``C''), and extended (``Ex'') -- defined in the following way:
\begin{itemize}
\item  Unresolved sources are sources in the high resolution image with 
$\theta_{M}<2$$\times$$\theta_{m}$.
\item Elongated sources are also sources in the high resolution image, but 
with $\theta_{M} \geq 2\times \theta_{m}$.  In practice, the major distinction 
between unresolved and elongated sources is that the major axis is resolved for
 elongated sources.
\item Complex sources are sources in the smoothed maps that overlap several 
sources in the high resolution maps.
\item Extended sources are sources in the smoothed maps which do not overlap 
sources in the high resolution maps. We defined ``ExG'' sources to be Extended 
sources detected in other radio surveys of M31.  Extended sources without 
counterparts in other radio surveys (``Ex'' sources) are {\it most likely 
false detections}, but are included here for completeness.
\end{itemize}
A total of 405 sources were detected: 281 are unresolved, 16 elongated, 51 
complex, and 57 extended -- of which two are ``ExG'' sources.  

The source flux density $S$ was calculated by correcting the value returned by
{\sc SFIND} for primary beam attenuation.  After the primary beam 
correction was applied, the errors in $\theta_{M}$, $\theta_m$, and 
$\theta_{PA}$ of each source were calculated using Equations 29a, 29b and 
30 in \citet{nvss}.  In this calculation, we assumed that the signal-to-noise 
of a source is equal to $S$/$\sigma_{RMS}$. We also calculated the error 
in integrated flux density ($S$) and peak brightness ($I$) using Equations 
36a and 37 in \citet{nvss} for a resolved (``C'', ``Ex'', and ``ExG'') source,
 Equations 39 and 41 in \citet{nvss} for a source where only the major axis is 
resolved (``El''), and Equation 43 in \citet{nvss} for an unresolved (``U'') 
source.  However, in all cases we set the ``clean bias'' term ($\sigma_{B}$ in 
the NVSS paper; \citeauthor{nvss} \citeyear{nvss}) to 0 because restrictive 
clean boxes were used.
 
To check the flux calibration, we compared the flux densities of GLG sources 
with the flux densities of their counterparts in the WENSS catalog 
\citep{wenss}, which was also conducted at 325 MHz. Figure \ref{glgwen} 
compares the WENSS and GLG flux density for all GLG sources with a ``good'' 
match in the WENSS catalog, determined using the criteria discussed below in 
Section \ref{specprop}.  In general, there is good agreement between 
the flux densities measured by the WENSS and GLG surveys, particularly 
for the strongest sources and unresolved sources -- implying that the flux 
calibration is correct.

After the flux errors were calculated, the positions of the GLG sources were 
registered to those of the strongest NVSS sources.  This was done by first 
finding all GLG sources that overlapped a NVSS source with 
$S_{1400}\geq15$~mJy, because the positions of these NVSS sources is known to 
better than 1\arcsec and match those in the optical reference frame to 
$\sim$40~mas \citep{nvss}.  Figure \ref{regpos} shows the offsets between the 
positions of these 15 GLG sources and their NVSS counterparts.  The average RA 
offset between the GLG and NVSS sources was -0\farcs099$\pm$0\farcs067 and the 
average DEC offset was 0\farcs115$\pm$0\farcs067, the error being the standard 
error in the mean.  Every source in the GLG source list was then shifted by 
the average offset, and the intrinsic uncertainty in RA ($\epsilon_{\alpha}$) 
and DEC ($\epsilon_{\delta}$) were set to the error of the offsets.  Errors in 
the positions of the GLG sources were then calculated with the formulae used 
for the NVSS source list \citep{nvss}.

\section{Source list}
\label{sourcelist}
The final source list is presented in Table \ref{srclist}.  The columns are
as follows:
\begin{description}
\item[Column 1:] Name of the source.  The sources are ordered in increasing 
distance from the center of the FOV.
\item[Column 2:] Type of source.  This is the morphological category of 
the sources as defined in Section \ref{srcprop}. ``U'' stands for unresolved, 
``El'' for elongated, ``C'' for complex, ``Ex'' for extended, and ``ExG'' for 
an Extended source with a counterpart in one or more other radio surveys of 
M31.
\item[Column 3:] Right Ascension of the source [J2000].\\
Error in RA ($\sigma_{\alpha}$), in arcseconds.
\item[Column 4:] Declination of the source [J2000].\\
Error in DEC ($\sigma_{\delta}$), in arcseconds.
\item[Column 5:] $\theta_{M31}$, angular distance from the center of M31, in 
arcmin. \\
$R_{M31}$, projected radius of source in M31 as defined in Section 
\ref{raddist}, in kpc.
\item[Column 6:] Major Axis ($\theta_M$) of the source, in arcseconds.\\
Error in Major Axis [$\sigma(\theta_M$)], in arcseconds.
\item[Column 7:] Minor Axis ($\theta_m$) of the source, in arcseconds.\\
Error in Minor Axis [$\sigma(\theta_m$)], in arcseconds.
\item[Column 8:] Position Angle ($\theta_{PA}$) N through E of the source, in 
degrees.\\
Error in Position Angle ($\sigma_{PA}$), in degrees.
\item[Column 9:] Peak brightness ($I$) of the source, in \mjyb.\\
Error in peak brightness ($\sigma_{I}$), in \mjyb.
\item[Column 10:] Integrated flux density ($S$) of the source in mJy.\\
Error in integrated flux density ($\sigma_{S}$), in mJy.
\item[Column 11:] RMS of the flux density around the source ($\sigma_{RMS}$), 
in mJy.
\item[Column 12:] Radio spectral index ($\alpha$) of source, as calculated in 
Section \ref{specprop}.\\
Spectral curvature parameter ($\varphi$), as calculated in Section 
\ref{specprop}.  The spectral indices of ``Ex'' sources without a counterpart 
in another radio survey were not published calculate because this source is most
 likely spurious, as mentioned in Section \ref{srcprop}.  It is also possible 
that some of the steeper spectrum ``U'' sources ($\alpha < -2$) are spurious 
as well.
\end{description}
 
\section{Properties of observed sources}
\label{props}
We analyzed the statistical properties of this population to see if radio 
sources in the field of M31 field differed from those elsewhere, and if so, 
how they differed.  We analyzed both the radial, flux density, and spectral 
distribution of the GLG sources, and for comparison, data from the WENSS 
\citep{wenss} and XMM-LSS radio survey \citep{aaron} surveys.  The WENSS 
survey mapped the entire sky north of $\delta=30^{\circ}$ at $\nu=325$~MHz 
with a limiting flux density density of 18 mJy and resolution of 
54\arcsec$\times$82\arcsec~around M31 \citep{wenss}, while the 325 MHz field 
of the XMM-LSS radio survey mapped a $\sim$5.6~$\deg^2$ region of sky with a 
resolution of 6.3\arcsec~and a limiting flux density of 4~\mjyb - properties 
similar to the observation presented here \citep{aaron}.

\subsection{Radial Distribution of Observed Sources}
\label{raddist}
If a substantial number of sources intrinsic to M31 were detected, we would 
expect to see an over-density of sources in the optical disk of M31.  In 
addition, any substructure present in M31 is most likely symmetric to some 
degree and might stand out in the radial distribution of sources.  
Unfortunately, interpreting the raw radial distribution of sources is difficult
 because the gain of the telescope decreases toward the edge of the FOV, as 
illustrated in Figure \ref{pbeam}.  As a result, faint sources detectable in 
the center of the FOV are undetectable at the edges, complicating models for 
the expected {\it observed} background radial distribution of sources.  
However, since the XMM-LSS 325~MHz observation was also done using one pointing
 of the VLA A-array, its primary beam shape is very similar to that of this 
observation.  Therefore, the effect of the primary beam's shape on the radial 
distribution of sources should be the same for these two datasets.  As a 
result, the XMM-LSS 325~MHz data provides a good background radial 
distribution for this comparison.  The XMM-LSS radio survey also observed a 
``blank'' field, so the radial distribution of the sources is dominated by 
instrumental effects and not by structure intrinsic to the program source.

We also compared the radial distribution of sources projected into the plane 
defined by the inclination of M31, which may contain more information about M31
 itself.  This was done by first converting the RA and DEC of each source in 
the GLG and XMM-LSS source list to angular coordinates ($X$,$Y$) relative to 
the center of M31 (or, for the XMM--LSS sources, the pointing center of 
that observation) using the {\sc IDL} task {\sc WCSSPH2XY}, and  then 
translating these angular 
coordinates into the plane of M31 on the sky ($X_{M31}$,$Y_{M31}$) using the 
following formulae:
\begin{eqnarray}
X_{M31}=X~\cos~52^{\circ}+Y~\sin~52^{\circ} \\
Y_{M31}=Y~\cos~52^{\circ}-X~\sin~52^{\circ}
\end{eqnarray}
where $52^{\circ}$ is the angle of the optical disk of M31 relative to N on the
 sky \citep{braun}.  Since the plane of M31 is inclined $77^{\circ}$ to the 
plane of the sky, a circle of radius R in M31 would be an ellipse in the 
($X_{M31}$,$Y_{M31}$) coordinate system.  Using this fact, the distance of a 
source from the center of the observation in the plane of M31 ($R_{M31}$) was 
calculated using the following formula:
\begin{equation}
R_{M31}=780~\mbox{kpc}~\times~\sqrt{X_{M31}^2 + 
\left(\frac{Y_{M31}}{cos~77^{\circ}}\right)^2.}
\end{equation}

As shown in Figure \ref{disdist}, there are substantial differences between the
 radial distribution of the XMM-LSS and GLG surveys.  The first difference is 
the number of sources in the XMM-LSS (256) and GLG survey (405).  Some of this 
difference is due to differences in the data analysis -- the XMM-LSS list does 
not include sources found in smoothed maps (``Ex'' GLG sources), only goes out 
to 1.34$^{\circ}$ from the pointing center (in the GLG survey, the furthest 
source is 1.75$^{\circ}$ from the pointing center), and does not include 
sources with $S_{325}<4$~mJy \citep{aaron} while the GLG survey has a 
5$\sigma$ sensitivity of $\sim$3~mJy.  However, after applying the above 
limitations to the GLG source list there are still significantly ($2.9\sigma$) 
more GLG sources (302) than XMM-LSS sources (256).\footnote{This implies a 
scaling factor of $\frac{302}{256}=1.18$ between the two surveys, used to 
generate the dot-dashed line in Figure \ref{disdist}.}  Some of this 
difference may be due to the fact that XMM-LSS source list was derived using 
the \aips~task {\sc VSAD} \citep{aaron}, a task similar to {\sc SFIND} without 
its FDR capabilities since the FDR algorithm allows {\sc SFIND} to detect
fainter sources than it would without it.

The second difference is in the distribution of sources between the GLG and 
XMM-LSS surveys -- both in the plane of the sky and in $R_{M31}$.  In the plane
 of the sky, the distribution of the XMM-LSS sources is relatively flat 
between 0.25$^{\circ}$ and 1.25$^{\circ}$, while the distribution of the GLG 
sources is rising in this range.  This difference is even more pronounced in 
the $R_{M31}$ distribution, where for $R_{M31}\ga 50$~kpc, there are noticeably
 more GLG sources than XMM--LSS sources.  Inside the optical disk of M31 
($R_{M31} \leq 27~\mbox{kpc}$), there are 145 GLG sources compared to
 130 XMM-LSS sources, a $\sim$0.5$\sigma$ excess.  If one applies the criteria 
of the XMM-LSS survey to the GLG source list, there are in fact fewer (119) 
GLG sources than XMM-LSS sources.  As a result, we expect that a majority of 
the GLG sources in the optical disk of M31 are background radio sources.

\subsection[]{Flux Distribution}
\label{flxdist}
If the 325~MHz population of M31 had a characteristic flux density, then one 
would see an over-density of GLG sources in a particular flux density range 
when compared with the flux distribution of the WENSS and XMM-LSS surveys.  As 
with the radial distribution, the primary beam affects the measured flux 
distribution, but only at low flux densities.  Therefore, we can compare the 
flux distribution of the GLG survey to that of both the WENSS and XMM-LSS 
surveys.  Figure \ref{flux} shows that the flux distribution of GLG sources 
does not have an over-density in a particular flux density bin or range of 
flux density bins -- implying that the flux distribution of sources in M31 is 
not substantially different from that of background sources.  However, given 
that we only expect to see a few sources in M31, they would need to be 
concentrated in one or two flux density bins for a difference to be detected.

In order to determine the completeness of the GLG survey, we compared the 
flux distribution with a model determined from the 327 MHz WSRT observations 
\citep{markw}.  As shown in Figure \ref{flux}, there is good agreement between 
the model and the observed flux distribution down to $\sim$3~mJy -- below the
5$\sigma$ sensitivity of the observations (4~mJy; \citeauthor{markw} 
\citeyear{markw}).  Therefore, we estimate that the GLG survey is complete to
between 3 and 4 mJy.

\subsection[]{Spectral Properties}
\label{specprop}
In general, the radio spectrum of a source can be expressed as a power law 
attenuated by some process.  In this paper, we parameterize the attenuation as 
if it were the result of free--free absorption because this the major cause of 
low--frequency spectral turnover in galactic objects.  Therefore, we fit the 
observed spectrum to the following function:
\begin{equation}
\label{specmodel}
S_{\nu}=S_{\nu_0}~\left(\frac{\nu}{\nu_0}\right)^{\alpha}~
e^{-\varphi(\frac{\nu}{1~GHz})^{-2.1}}
\end{equation}
where $S_{\nu}$~is the flux density of the source at a frequency $\nu$,
$S_{\nu_0}$~is the flux density of the source at frequency $\nu_0=324.5$~MHz, 
$\alpha$ is the radio spectral index, and $\varphi$ is a parameter that 
measures spectral curvature.  If the spectral curvature is truly the result of 
free--free absorption, then $\varphi>0$ and corresponds to the optical depth 
due to free-free absorption at $\nu=1~$GHz, equal to \citep{condon}:
\begin{equation}
\varphi=3.3\times10^{-7}(EM)\left(\frac{T_e}{10^4~K}\right)^{-1.35}
\end{equation}
\begin{equation}
\label{em}
EM=\int n_e^2 dl
\end{equation}
where $EM$ is the Emission Measure in units of cm$^{-6}$~pc, $T_e$ is the 
electron temperature (in all $EM$ calculations presented here we assume
 $T_e=10^4$~K), $n_e$ is the electron density in units of cm$^{-3}$, and the 
integral in Equation (\ref{em}) extends over the line-of-sight to the object.
  For sources with an insufficient number of counterparts ($n_{ctrps}\leq3$) at
other radio frequencies, the model used does not include the attenuation term 
and the spectrum was fit to a simple power law.

The surveys used to calculate the radio spectral index ($\alpha$) and spectral 
turnover ($\varphi$) of the GLG sources were the 36W ($\nu=610$~MHz;~
\citeauthor{36w} \citeyear{36w}), 37W ($\nu=1412$~MHz;~\citeauthor{37w} 
\citeyear{37w}), Braun 1990 [Braun] ($\nu=1465$~MHz;~\citeauthor{braun} 
\citeyear{braun}), and NVSS ($\nu=1400$~MHz;~\citeauthor{nvss} \citeyear{nvss})
 surveys. The properties of these surveys are shown in Table \ref{catprop}.  If
 a source in another catalog was less than:
\begin{equation}
\label{radctrp}
r=Max\left(\frac{\theta_M^{GLG}+\theta_m^{GLG}}{2},\frac{\theta_M+\theta_m}{2}\right)
\end{equation} 
away from a GLG source they were classified as a match, where $\theta_M^{GLG}$ 
is the major axis of the GLG source, $\theta_m^{GLG}$ is the minor axis of the 
GLG source (these are not the deconvolved values but the size of the source in 
our images), $\theta_M$ is the deconvolved major axis of the catalog source, 
and $\theta_m$ is the deconvolved minor axis of the catalog source 
\citep{braun}.\footnote{Because the 36W and 37W surveys were done with the 
Westerbork Radio Telescope (WSRT) as opposed to the VLA, there is a systematic 
offset in position between them and the GLG survey.  However, this difference 
is $<$1\arcsec, much smaller than $>$6\arcsec~radius used to search for 
counterparts.}   A ``good match'' between a GLG source and a source in one of 
these catalogs was when the GLG had only one match in the other catalog and its
 counterpart only had no other matches in the GLG source list.  The results of 
this comparison are summarized in Table \ref{radcomp}.  The number of false 
matches was calculated by shifting the position of the radio catalog in 
question by every combination of $\pm$1\arcmin,0\arcmin~in RA and 
$\pm$1\arcmin,0\arcmin~in DEC (except for a shift of [0,0]), and calculating 
the number of matches and good matches in each combination.  The number of 
false (good) counterparts reported in Table \ref{radcomp} is the average number
 of (good) matches detected in the eight shifted datasets, and the error in 
these values is the standard deviation of the number of (good) matches.  As 
seen in Table \ref{radcomp}, the number of false counterparts and false good 
counterparts are much less that observed.  The WENSS survey has a 
larger number of false comparisons than the other surveys because its beam is 
much larger, resulting in a larger search radius.  In order to avoid problems 
with source confusion, $\alpha$ and $\varphi$ were calculated using only values
 from a ``good match'' by fitting the spectrum to the form defined in Equation 
(\ref{specmodel}) using a least-mean-squares algorithm.

For GLG sources without a good counterpart in any of these radio surveys, we 
calculated the upper limit on the spectral index $\alpha$.  In order to do 
this, we made a sensitivity map\footnote{A sensitivity map shows the RMS of an
observation over the field-of-view} of the 36W, 37W, and Braun surveys (shown 
in Figure \ref{pbeam}) by first making a sensitivity map of each pointing in 
the survey and then combining them together, weighing each pointing by 
1/$\sigma^2$ where $\sigma$ is the RMS of the pointing map.  The upper limit 
was then determined by adjusting the stated minimum detectable flux density by 
the ratio of the survey's sensitivity at the location of the source to the 
survey's peak sensitivity.  This was not necessary for the NVSS survey because 
our FOV is entirely within theirs, and 
an upper--limit of $S_{1400}=$15~mJy (the completeness limit of the NVSS 
survey) was used \citep{nvss}.  The spectral index of these sources given in 
Table \ref{srclist} is the flattest spectrum consistent with all 
non-detections.  The spectral energy diagram (SEDs) of all GLG sources can be 
found in Figure \ref{sed}, and the calculated $\alpha$ and $\varphi$ of each 
source can be found in Table \ref{srclist}.

For sources with a ``good match'' in the 36W (610~MHz) survey and in at least 
one of the 1400~MHz surveys (NVSS, 37W, or Braun), it was possible to compare 
their $\alpha_{325}^{610}$ and $\alpha_{325}^{1400}$ spectral indices, the 
spectral index between 325 and 610 MHz and 325 and 1400 MHz respectively.  For 
sources with counterparts in multiple 1400~MHz radio catalogs, 
$\alpha_{325}^{1400}$ is the weighted average of the individual spectral 
indices.  Figure \ref{colcol} shows the graph of $\alpha_{325}^{1400}$ vs. 
$\alpha_{325}^{610}$ for the 156 GLG sources that met this criteria.  These 
sources are not evenly distributed around the 
$\alpha_{325}^{1400}=\alpha_{325}^{610}$ line; 99 GLG sources have 
$\alpha_{325}^{610}>\alpha_{325}^{1400}$ while only 57 have 
$\alpha_{325}^{610} \leq \alpha_{325}^{1400}$.  A larger number of 
$\alpha_{325}^{610}>\alpha_{325}^{1400}$ sources is expected because the 
synthesized beam of the 36W survey is larger than that of any of the 1400~MHz 
surveys, meaning that the reported flux density of 36W sources possibly 
include emission resolved out in the 1400~MHz surveys.  However, 23 of these 
sources have $\mid\alpha_{325}^{610}-\alpha_{325}^{1400}\mid > 3 \times 
Max(\sigma_{\alpha_{325}^{610}},\sigma_{\alpha_{325}^{1400}})$, where 
$\sigma_{\alpha_{325}^{610}}$ is the error of $\alpha_{325}^{610}$ and  
$\sigma_{\alpha_{325}^{1400}}$ is the error of $\alpha_{325}^{1400}$, a 
statistically significant difference that cannot be explained by a difference 
in beam size.  All but one of these sources have 
$\alpha_{325}^{610}>\alpha_{325}^{1400}$ -- the signature of both free-free 
absorption and synchrotron self--absorption, and such sources have been 
observed in previous radio surveys, e.g. \citet{hzrg}.  GLG186, the only such 
source with
 $\alpha_{325}^{610}<\alpha_{325}^{1400}$ -- has been categorized a variable 
source (see Paper II \citep{paper2} for more information).

To determine if the GLG sources -- both in and out of the optical disk of M31 
-- had different spectral properties than background sources, we compared the 
$S_{325}$-$\alpha$ distribution \citep{zhang} and the $\alpha$ number 
distribution (Figures 7 and 8 in \citeauthor{hzrg} \citeyear{hzrg}) of the 
GLG sources -- shown in Figure \ref{alpgraph} -- with that of sources in both 
the WENSS and NVSS survey.  Both the $S_{325}$-$\alpha$ and 
$\alpha$ number distribution of GLG sources is consistent with the WENSS/NVSS 
results.  We also compared the spectral index distribution of GLG sources with 
a higher-frequency radio counterpart inside and outside the optical disk of M31
 to see if there are any differences between these populations, also shown in 
Figure \ref{alpgraph}.  While there are substantial selection effects involved,
 e.g. the FOV of the deeper 1.4 GHz surveys are 
concentrated on the optical disk of M31 so steep sources outside M31 might be 
missed, there is no noticeable difference in the number distribution of 
$\alpha$ between these regions.  However, this is not unexpected since the 
dominant radio sources of a galaxy -- SNRs and HII regions -- have similar 
spectral indices distributions than that of the dominant forms of background 
sources -- FRI and FRII radio galaxies.

\section{Conclusions}
\label{conclusions}
In this paper, we present the results of a high resolution (6\arcsec) 325 MHz 
survey of a 7.6 deg$^2$ field centered on M31.  In this dataset, we identified 
405 discrete 
sources (the ``GLG'' source list), and determined their size, orientation, and 
flux density.  Through comparisons with 610 and 1400 MHz radio surveys of the 
same region, we calculated the spectral index $\alpha$ and spectral curvature 
$\varphi$ of these sources.

To determine the nature of the GLG source list, we compared its properties to 
that of source lists derived from blank-field and/or large-area radio surveys 
-- the WENSS, NVSS, and XMM-LSS radio surveys.  A comparison of the radial 
distribution of GLG and XMM-LSS 
radio sources - both in the plane of the sky and in the plane of M31 - showed 
that there is not a significant concentration of sources in the optical 
disk of M31.  However, there does appear to be an excess of GLG sources at the 
edge of the field-of-view.  The reason for this is unknown, but most likely due
to the different source finding programs used between the surveys.  A 
comparison between the flux distribution of GLG sources with that of the WENSS 
and XMM-LSS radio surveys revealed no statistically significant over-density or
 under-densities in a sizable flux range.  The $\alpha$ number distribution of
 GLG sources was similar to that of sources in both the WENSS and NVSS radio 
surveys, as 
was the relationship between 325 MHz flux density $S_{325}$ and $\alpha$.  
The spectral index distribution of sources within the optical disk of M31 was
also similar to that of sources in outside the optical disk of M31.  All of 
this implies that the GLG source list is dominated by background radio galaxies
 and not sources intrinsic to M31.

In an accompanying paper \citep{paper2}, we compare the GLG source list with 
that of far-IR, IR, optical, and X--ray catalogs of this region and use 
this multi-wavelength information, as well as the morphology and spectral 
index/curvature of a source, to determine the nature of individual sources.  
Through the method described in that paper, we have identified five supernova 
remnant and three pulsar wind nebulae candidates in M31, as well as a galaxy 
merger, BL Lac candidate, Giant Radio Galaxy candidate, and several 
low-frequency variable and ultra-steep spectrum ($\alpha < -1.6$) sources.

\acknowledgments
 The authors thank Andrew Hopkins for providing us with the latest version of 
{\sc SFIND}.  TJW Lazio acknowledges that basic research in radio astronomy at 
the NRL is supported by the Office of Naval Research.   JDG thanks Elias 
Brinks, Mike Garcia, Phil Kaaret, Albert Kong, Linda Schmidtobreick, Sergey 
Trudolyubov, Rene Walterbos, Ben Williams, C. Kevin Xu, and the SIMBAD help 
desk for providing us with source lists and/or images; Carole Jackson for 
providing her model of the flux distribution of background radio galaxies at 
325~MHz; and Aaron Cohen, Elly Berkhuijsen, Jim Cordes, Rosanne DiStefano, 
Mike Garcia, Dan Harris, John Huchra, Namir Kassim, Albert Kong, Pat Slane, 
Krzysztof Stanek, Lorant Sjouwerman, Ben Williams, and Josh Winn for useful 
discussions; and Harvard CDF for computer access.

The National Radio Astronomy Observatory is a facility of the National Science 
Foundation operated under cooperative agreement by Associated Universities, 
Inc. This research has made use of NASA's Astrophysics Data System; of the 
SIMBAD 
database, operated at CDS, Strasbourg, France; of the NASA/IPAC Extragalactic 
Database (NED) which is operated by the Jet Propulsion Laboratory, California 
Institute of Technology, under contract with the National Aeronautics and 
Space Administration; of data products from the Two Micron All Sky Survey, 
which is a joint project of the University of Massachusetts and the Infrared 
Processing and Analysis Center/California Institute of Technology, funded by 
the National Aeronautics and Space Administration and the National Science 
Foundation; and of the NASA/ IPAC Infrared Science Archive, which is operated 
by the Jet Propulsion Laboratory, California Institute of Technology, under 
contract with the National Aeronautics and Space Administration.
\clearpage
\bibliography{ms}
\bibliographystyle{apj}
\clearpage
\begin{deluxetable}{c|cccccc|c}
\tablecolumns{8}
\tablewidth{0pc}
\tableheadfrac{0.05}
\tabletypesize{\scriptsize}
\tablecaption{Summary of Radio Surveys \label{catprop}}
\tablehead{
\colhead{Survey} & \colhead{Observation} & \colhead{Frequency} & 
\colhead{Flux Density} & \colhead{Resolution} & \colhead{FOV} & 
\colhead{\# of} & \colhead{Reference} \\
\colhead{} & \colhead{Date} & \colhead{} & \colhead{Limit\tablenotemark{a}} & 
\colhead{} & \colhead{} & \colhead{Sources} & \colhead{} \\
\colhead{} & \colhead{} & \colhead{[MHz]} & \colhead{[mJy]} & 
\colhead{[\arcsec]} & \colhead{[deg$^2$]} & \colhead{} & \colhead{} }
\startdata
WENSS & 1991--1993 & 325 & 18 & 54$\times$82 & \nodata & 288 & \cite{wenss}\\
36W & 1975 & 610 & 3 & 54$\times$82 & 10.1 & 392 & \cite{36w} \\
37W & 1975--1979 & 1412 & 1 & 23$\times$35 & 3.1 & 249 & \cite{37w} \\
Braun & 1990 & 1465 & 0.15 & 5 & 1.9 & 534 & \cite{braun} \\
NVSS & 1993--1996 & 1400 & 2.5 & 45 & \nodata & 831 & \cite{nvss} \\
\hline
GLG & 2000 & 325 & 3 & 6 & 7.6 & 405 & This paper \\
\enddata
\tablenotetext{a}{The Flux Density Limits given for WENSS, 36W, 37W, Braun, 
and NVSS surveys are the lowest flux densities quoted by these authors.  For 
all  surveys but the 36W survey, this corresponds to a 5$\sigma$ cutoff.
For the 36W survey, this correspond to a 4$\sigma$ cutoff.}
\end{deluxetable}

\begin{table}
\caption{{\bf Observation Properties} \label{obsprop}}
\begin{center}
\begin{tabular}{ccc}
\hline
\hline
\multicolumn{3}{c}{{\it Observed Sources}} \\
\hline
Source & Integration Time [Minutes] & Number of Visibilities \\
3C405 & 25.8 & 47330 \\
TXS 0035+413 & 23.1 & 42250 \\
M31 & 239.5 & 438657 \\
\hline
\multicolumn{3}{c}{{\it Receiver Properties}} \\
\hline
IF Number & Frequency [MHz] & Bandwidth (Channel Separation) [kHz] \\
1 & 327.50 & 3125.0 (48.83) \\
2 & 321.56 & 3125.0 (48.83) \\
\hline
\multicolumn{3}{c}{{\it Primary Beam properties}} \\
\hline
Full Width at Half Maximum & \multicolumn{2}{c}{138\arcmin} \\ 
Field of View & \multicolumn{2}{c}{7.6~$\deg^2$} \\
Resolution & \multicolumn{2}{c}{6\arcsec$\times$6\arcsec} \\
RMS at center of Field & \multicolumn{2}{c}{0.61~mJy} \\
\hline
\multicolumn{3}{c}{{\it General Properties}} \\
\hline
Date of Observation & \multicolumn{2}{c}{2000 DEC 15} \\
Pointing Center [J2000] &  RA: $00^{\mbox{h}}~42^{\mbox{m}}~45\fs9$ &  DEC: 
$41^{\circ}~16^{\prime}~18\farcs0$ \\
\hline
\hline
\end{tabular}
\end{center}
\end{table}

\begin{deluxetable}{cc||rr|c||ccc|ccc||c}
\tablecolumns{12}
\tablewidth{0pc}
\tableheadfrac{0.05}
\tabletypesize{\scriptsize}
\tablecaption{GLG Source list \label{srclist}}
\setlength{\tabcolsep}{0.02in}
\tablehead{
\colhead{Name} & \colhead{Type} & \colhead{RA [h m s]} & 
\colhead{DEC [$^{\circ}~^{\prime}~^{\prime \prime}$]} & 
\colhead{$\theta_{M31}$~[\arcmin]} & 
\colhead{$\theta_{M}$~[\arcsec]} & 
\colhead{$\theta_{m}$~[\arcsec]} & 
\colhead{$\theta_{PA}$~[$^{\circ}$]} & \colhead{$I$~[$\frac{mJy}{beam}$]} & 
\colhead{$S$~[mJy]} & 
\colhead{$\sigma_{RMS}$} & \colhead{$\alpha$} \\
\colhead{} & \colhead {} & 
\colhead{$\sigma_{\alpha}$~[\arcsec]} & 
\colhead{$\sigma_{\delta}$~[\arcsec]} & 
\colhead{$R_{M31}$~[kpc]} & 
\colhead{$\sigma(\theta_M)$} & 
\colhead{$\sigma(\theta_m)$} & 
\colhead{$\sigma_{PA}$} & \colhead{$\sigma_A$} & 
\colhead{$\sigma_S$} & \colhead{} & 
\colhead{$\varphi$}} 
\startdata
GLG001 &  U & 00~42~50.40 & +41~16~10.28 &  0.8 &  5.7 &  4.6 &  15 &  7.94 & 
 8.26 &  0.69 & -0.64$\pm$0.06 \\*
\nodata & \nodata & 0.34 & 0.40 &  0.7 & 0.4 & 0.3 & 6 & 0.70 & 0.70
 & \nodata & \nodata \\
GLG002 &  U & 00~42~26.62 & +41~18~04.62 &  4.0 &  4.9 &  4.5 &  76 &  4.17 & 
 3.65 &  0.63 & -1.45$\pm$0.34 \\*
\nodata & \nodata & 0.71 & 0.66 &  3.9 & 1.4 & 1.2 & 155 & 0.59 & 0.59
 & \nodata & 0.13$\pm$0.05 \\
GLG003 &  U & 00~42~32.90 & +41~11~43.57 &  5.1 &  5.0 &  4.7 &  14 &  5.12 & 
 4.76 &  0.54 & -0.95$\pm$0.02 \\*
\nodata & \nodata & 0.46 & 0.48 &  1.4 & 0.6 & 0.5 & 26 & 0.53 & 0.53
 & \nodata & \nodata \\
GLG004 &  U & 00~43~18.00 & +41~17~21.94 &  6.1 &  10.0 &  7.7 &  67 &  7.37
 &  22.85 &  0.60 & -0.84$\pm$0.14 \\*
\nodata & \nodata & 0.22 & 0.19 &  4.2 & 0.1 & 0.0 & 6 & 0.70 & 0.70
 & \nodata & 0.04$\pm$0.02 \\
\enddata
\end{deluxetable}

\begin{deluxetable}{cc|cc|cc|c}
\tablecolumns{7}
\tablewidth{0pc}
\tableheadfrac{0.05}
\tabletypesize{\scriptsize}
\tablecaption{Comparison with other Radio Surveys \label{radcomp}} 
\setlength{\tabcolsep}{0.02in}
\tablehead{
\colhead{Survey} & \colhead{\# of GLG sources} & 
\colhead{\# of GLG sources} & 
\colhead{\# of GLG sources} & \colhead{\# of false} & 
\colhead{\# of false} & \colhead{Reference} \\
\colhead{} & \colhead{in survey FOV} & \colhead{with counterparts}
 & \colhead{with good counterparts} & \colhead{counterparts} & 
\colhead{good counterparts}
}
\startdata
WENSS & 405 & 121 & 89 & 35.4$\pm$8.2 & 16.3$\pm$4.5 & \cite{wenss} \\
36W & 357 & 176 & 161 & 8.8$\pm$2.1 & 4.3$\pm$1.8 & \cite{36w} \\
37W & 227 & 153 & 145 & 1.8$\pm$1.6 & 1.4$\pm$1.3 & \cite{37w} \\
Braun & 87 & 67 & 53 & 9.8$\pm$2.6 & 4.9$\pm$2.4 & \cite{braun} \\
NVSS & 405 & 189 & 185 & 2.3$\pm$1.6 & 2.1$\pm$1.6 & \cite{nvss} 
\enddata
\end{deluxetable}

\clearpage
\begin{center}
\begin{minipage}{12cm}
\end{minipage}\    \
\begin{minipage}{12cm}
\end{minipage}\\
\begin{minipage}{12cm}
\end{minipage}\    \
\begin{minipage}{12cm}
\end{minipage}
\end{center}
\figcaption[]{Contours showing the 1$\sigma$ sensitivity of the ({\it clockwise
 from upper left:}) GLG (this paper), 36W \citep{36w}, Braun \citep{braun}, and
 37W \citep{37w} radio surveys across the field of M31.  The sensitivity for 
the 36W, Braun, and 37W surveys were generated using the procedure described 
in Section \ref{specprop}.  The optical image is of M31 from the Palomar 
Optical Sky Survey.  For the GLG and 36W sensitivity maps, the contour levels 
are 0.6, 0.65, 0.7,...,1.0, 1.25, 1.5, ..., 3.0 \mjyb.  The jagged edges of the
 GLG sensitivity map 
are due to the shapes of the facets discussed in Section \ref{imgprod}. For the
 37W sensitivity map, the contour levels are 0.1, 0.15, 0.2,..., 1.0 \mjyb, and
 for the Braun sensitivity map, the contour levels are 0.025, 0.035, 0.045, 
0.05, 0.06,..., 0.1, 0.25 \mjyb.\label{pbeam}}

\newpage
\begin{center}
\begin{minipage}{20cm}
\end{minipage}
\figcaption{Flow chart of the data reduction described in Sections 
\ref{datared}, \ref{imgprod}, and \ref{gensrc}.\label{flwchrt}}
\end{center}

\newpage
\begin{center}
\begin{minipage}{15cm}
\end{minipage}
\end{center}
\figcaption{Palomar Optical Sky Survey image of M31 overlaid with the position 
of the GLG sources (white contours representing the major and minor axis of 
the sources given in Table \ref{srclist}) and the 1$\sigma$ sensitivity of the 
GLG survey (black contours).  The sensitivity contours correspond to 
0.6, 0.65, 0.7, 0.75, 0.8, 0.85, 0.9, 0.95, 1.0, 1.5, 2.0, 2.5 \mjyb. 
\label{m31_325}}

\newpage
\begin{center}
\begin{minipage}{15cm}
\end{minipage}
\end{center}
\figcaption{Flux Density of WENSS sources with a ``good'' match (as defined in 
Section \ref{specprop}) in the GLG catalog vs. flux density of corresponding 
GLG sources.  The diamonds indicate an ``U'' or ``El'' GLG source, and the 
sloping straight-line indicates where the WENSS flux density equals the 
GLG~flux density.\label{glgwen}}

\newpage
\begin{center}
\begin{minipage}{15cm}
\end{minipage}
\end{center}
\figcaption{Positional offset (GLG-NVSS) between the 15 NVSS and GLG sources 
with $S_{1400}>15$~mJy.  These sources were used to register the location of 
the GLG sources.\label{regpos}}

\newpage
\begin{center}
\begin{minipage}{15cm}
\end{minipage}\\
\begin{minipage}{15cm}
\end{minipage}
\end{center}
\figcaption[]{Radial distribution ({\it top}) and $R_{M31}$ distribution 
({\it bottom}; defined in Section \ref{raddist}) of GLG (filled circles) 
and XMM--LSS survey (solid line) sources.  The diamonds indicate all 
non-extended GLG sources, while the triangles are non-extended GLG sources with
 $S_{325}>4$~mJy within 1.34$^\circ$ of the pointing center, i.e. the limits of
 the XMM--LSS survey \citep{aaron}.  The dot-dashed line is the distribution of
 the XMM--LSS survey scaled by the ratio of ``triangle'' GLG sources to XMM-LSS
 sources (see Section \ref{raddist} for details).  The black vertical line in 
the bottom graph marks the outer edge of the optical disk of 
M31.\label{disdist}}

\newpage
\begin{center}
\begin{minipage}{15cm}
\end{minipage}\     \
\begin{minipage}{15cm}
\end{minipage}
\end{center}
\figcaption[]{{\it Top:} The flux distribution of the GLG survey (plus signs),
 WENSS 
survey (solid line), and the XMM-LSS survey
(dot-dashed line).  The y-axis is $S_{\nu}^{\frac{5}{2}}n(S)$ where 
$n(S)=\frac{N(S_0<S<S_1)}{S_1-S_0} \div FOV$ and $N(S_0<S<S_1)$ is the total 
number of sources with flux density S between $S_0$ and $S_1$.  The 
$S_{\nu}^{\frac{5}{2}}$ term corrects for Euclidean geometry \citep{nvss}. The 
diamonds represent non-extended sources.  The vertical lines represent the 
completeness limit of the XMM-LSS (left; \citeauthor{aaron} \citeyear{aaron}) 
and WENSS (right; \citeauthor{wenss} \citeyear{wenss}) surveys.\\
{\it Bottom:} The flux distribution of the GLG survey (plus signs) compared
with the 327 MHz flux distribution model (line) derived from WSRT observations 
\citep{markw}. \label{flux}}

\newpage

\begin{center}
\begin{minipage}{15cm}
\end{minipage}
\end{center}
\figcaption[]{Radio Spectrum ($\nu=325-1400$~MHz) of all GLG sources.  A 
filled circle represents a good match, an arrow represents an upper--limit, an 
open square on the x--axis represents a ``bad match'' (the GLG source either 
had more than one counterpart in this catalog or the catalog source had more 
than one counterpart in the GLG source list), and the dashed line is the best 
fit to Equation \ref{specmodel}.  The upper-limits are scaled by the local rms
of the survey in question.\label{sed}}

\newpage
\begin{center}
\begin{minipage}{15cm}
\end{minipage}
\end{center}
\figcaption{``Color-Color'' diagram of all 156 GLG sources with a good match 
in the 36W catalog and at least one of the 1400 MHz catalogs.  The diamond is 
the one extended source that met this criteria, and the squares are sources 
with $\mid\alpha_{325}^{610}-\alpha_{325}^{1400}\mid>3 
\times Max(\sigma_{\alpha_{325}^{610}},\sigma_{\alpha_{325}^{1400}})$. 
\label{colcol}}

\newpage
\begin{center} 
\begin{minipage}{10cm}
\end{minipage}\    \
\begin{minipage}{10cm}
\end{minipage}\    \
\begin{minipage}{10cm}
\end{minipage}\    \
\begin{minipage}{10cm}
\end{minipage}
\end{center}
\figcaption[]{{\it Clockwise from top left:} $\alpha$ vs. $S_{325}$ for the 
244 GLG sources with a good counterpart in at least one other radio catalog; 
median spectral index vs. 325~MHz flux density for these sources (data points 
are a
binned version of the top left graph, and the error bars are the standard 
deviation of spectral index in the $S_{325}$ bin) overlaid with the results 
from the WENSS--NVSS surveys (triangles; \citet{zhang}); number distribution 
of $\alpha$, and fractional distribution of $\alpha$ for sources inside 
(solid line) and outside (dashed line) the optical disk of M31.
\label{alpgraph}}

\end{document}